# Biomarker Investigation using Multiple Brain Measures from MRI through XAI in Alzheimer's Disease Classification

D. Coluzzi [1,*], V. Bordin[1,*], M. W. Rivolta[2], I. Fortel[3], L. Zhang[4], A. Leow[3, 5], G. Baselli[1]

[1] Dipartimento di Elettronica, Informazione e Bioingegneria, Politecnico di Milano, Milan, Italy
[2] Dipartimento di Informatica, Università degli Studi di Milano, Milan, Italy
[3] Department of Biomedical Engineering, University of Illinois Chicago, Chicago, IL, USA
[4] Department of Electrical and Computer Engineering, University of Pittsburgh, Pittsburgh, PA, USA
[5] Department of Psychiatry, University of Illinois Chicago, Chicago, IL, USA

\* Equal contributions

\# Corresponding author:
Davide Coluzzi
Politecnico di Milano
Department of Electronics, Information and Bioengineering
Via Camillo Golgi 39, 20133 Milan
davide.coluzzi@polimi.it

**ABSTRACT**

Alzheimer's Disease (AD) is the world leading cause of dementia, a progressively impairing condition leading to high hospitalization rates and mortality. To optimize the diagnostic process, numerous efforts have been directed towards the development of deep learning approaches (DL) for the automatic AD classification. However, their typical black box outline has led to low trust and scarce usage within clinical frameworks. In this work, we propose two state-of-the art DL models, trained respectively on structural MRI (ResNet18) and brain connectivity matrixes (BC-GCN-SE) derived from diffusion data. The models were initially evaluated in terms of classification accuracy. Then, results were analyzed using an Explainable Artificial Intelligence (XAI) approach (Grad-CAM) to measure the level of interpretability of both models. The XAI assessment was conducted across 132 brain parcels, extracted from a combination of the Harvard-Oxford and AAL brain atlases, and compared to well-known pathological regions to measure adherence to domain knowledge. Results highlighted acceptable classification performance as compared to the existing literature (ResNet18: $TPR_{median}$ = 0.817, $TNR_{median}$ = 0.816; BC-GCN-SE: $TPR_{median}$ = 0.703, $TNR_{median}$ = 0.738). As evaluated through a statistical test ($p < 0.05$) and ranking of the most relevant parcels (first 15%), Grad-CAM revealed the involvement of target brain areas for both the ResNet18 and BC-GCN-SE models: the medial temporal lobe and the default mode network. The obtained interpretabilities were not without limitations. Nevertheless, results suggested that combining different imaging modalities may result in increased classification performance and model reliability. This could potentially boost the confidence laid in DL models and favor their wide applicability as aid diagnostic tools.

**Table of Acronyms**

| FP | Frontal Pole | pTFusC | Temporal Fusiform Cortex, posterior division | CaN | Caudate |
| SFG | Superior Frontal Gyrus | CO | Central Opercular Cortex | Pu | Putamen |
| MidFG | Middle Frontal Gyrus | PP | Planum Polare | Pal | Pallidum |
| IFG_tri | Inferior Frontal Gyrus, pars triangularis | HG | Heschls Gyrus | Hip | Hippocampus |
| IFG_oper | Inferior Frontal Gyrus, pars opercularis | PT | Planum Temporale | Amg | Amygdala |
| PreCG | Precentral Gyrus | PostCG | Postcentral Gyrus | NAcc | Accumbens |
| SMA | Juxtapositional Lobule Cortex - formerly Supplementary Motor Cortex | SPL | Superior Parietal Lobule | Cereb1 | Cerebelum Crus1 |
| FOrb | Frontal Orbital Cortex | aSMG | Supramarginal Gyrus, anterior division | Cereb2 | Cerebelum Crus2 |
| FO | Frontal Operculum Cortex | pSMG | Supramarginal Gyrus, posterior division | Cereb3 | Cerebelum 3 |
| MedFC | Frontal Medial Cortex | AG | Angular Gyrus | Cereb45 | Cerebelum 4 5 |
| IC | Insular Cortex | PO | Parietal Operculum Cortex | Cereb6 | Cerebelum 6 |
| AC | Cingulate Gyrus, anterior division | PrCun | Precuneous Cortex | Cereb7 | Cerebelum 7b |
| PC | Cingulate Gyrus, posterior division | sLOC | Lateral Occipital Cortex, superior division | Cereb8 | Cerebelum 8 |
| SubCalC | Subcallosal Cortex | iLOC | Lateral Occipital Cortex, inferior division | Cereb9 | Cerebelum 9 |
| PaCiG | Paracingulate Gyrus | ICC | Intracalcarine Cortex | Cereb10 | Cerebelum 10 |
| TP | Temporal Pole | Cuneal | Cuneal Cortex | Ver12 | Vermis 1 2 |
| aSTG | Superior Temporal Gyrus, anterior division | aPaHC | Parahippocampal Gyrus, anterior division | Ver3 | Vermis 3 |
| pSTG | Superior Temporal Gyrus, posterior division | pPaHC | Parahippocampal Gyrus, posterior division | Ver45 | Vermis 4 5 |
| aMTG | Middle Temporal Gyrus, anterior division | LG | Lingual Gyrus | Ver6 | Vermis 6 |
| pMTG | Middle Temporal Gyrus, posterior division | TOFusC | Temporal Occipital Fusiform Cortex | Ver7 | Vermis 7 |
| toMTG | Middle Temporal Gyrus, temporooccipital part | OFusG | Occipital Fusiform Gyrus | Ver8 | Vermis 8 |
| aITG | Inferior Temporal Gyrus, anterior division | SCC | Supracalcarine Cortex | Ver9 | Vermis 9 |
| pITG | Inferior Temporal Gyrus, posterior division | OP | Occipital Pole | Ver10 | Vermis 10 |
| toITG | Inferior Temporal Gyrus, temporooccipital part | Tha | Thalamus | BSt | Brain-Stem |
| aTFusC | Temporal Fusiform Cortex, anterior division | | | | |

*Brain parcel acronyms of the Harvard-Oxford combined with AAL atlas.*

## 1. INTRODUCTION

As populations continue to age, dementia cases are on the rise, which pose a serious public health risk and place a huge social and economic burden on many countries. The most prevalent cause of dementia is Alzheimer's Disease (AD), a neurodegenerative pathology occurring when nerve cells die in the brain, which initially manifests with impaired memory [1]. Dementia is a common disorder among the elderly population which is associated with significant disability, increased hospitalization, and mortality. In recent times, research focused extensively on AD and neuroimaging techniques emerged as a critical tool for diagnosing and monitoring the disease progressions. Structural brain changes, such as those visible on magnetic resonance imaging (MRI), can reveal the early involvement of the medial temporal lobe (MTL) in AD, particularly in terms of hippocampal and parahippocampal atrophy [2], [3]. Additionally, numerous studies linked the entorhinal cortex to changes in the cognitive performance of diseased individuals [4]. The brain cortex is also involved in the neurodegenerative process underlying AD. Both temporal and parietal regions present atrophic changes in correspondence of the early stages of the condition [5]. On the other hand, the symptomatic progression is often accompanied by a more extensive brain cortical thinning alongside ventricular enlargement [2], [6].

In addition, AD assessment can be supplemented with complementary information through diffusion MRI (dMRI) and fiber tracking due to dendritic, myelin and axonal loss which accompanies atrophy [4]. From dMRI measures, it is possible to extract structural connectivity data, depicting the brain as a graph, parcels as nodes, and derived dMRI metrics (such as the number of reconstructed streamlines) as edges. Coherent changes across different dMRI metrics used for edge-weighting in the connectivity structure due to AD were widely documented, occurring across a range of spatial scales and levels [7]. At the global level, AD was depicted as a disconnection syndrome which can be characterized by the connectome degeneration, affecting graph topology, that is governed by long-range connections [8], [9]. At sub-graph level, studies demonstrated that the Default Mode Network (DMN), which is involved in memory processes, is vulnerable to atrophy, amyloid protein deposition, and white matter microstructure alterations resulting in a disrupted structural connectivity configuration [7], [10], [11]. AD abnormalities in structural connectivity and dMRI studies also found the temporal lobe, whose disruptions contribute to memory impairment [7], [12], [13], and some regions of the MTL [14], which are also often reported in the DMN. These findings were corroborated by functional connectivity studies which consistently revealed a decreased functional connectivity between the posterior and anterior portions of the DMN [15], [16]. DMN in general was found to be replicable hallmark and an important area for AD in either structural and functional connectivity analysis, highlighting many common regions [17].

With the growing global incidence of neurodegenerative disorders, such as AD, a heightened interest to advance in areas such as diagnosis, treatment, prevention, drug discovery, and provision of improved healthcare services was noticed. Moreover, clinical decision support systems are possible to be developed through AI methods. Most of the research in the domain of AD, and more in general, neurodegeneration focused on using brain imaging. Traditional Machine Learning methods with well-known classification methods like Logistic Regression, Random Forest and Support Vector Machines were widely utilized both from MRI (mainly T1-weighted, but also Fluid Attenuated Inversion Recovery - FLAIR and T2-weighted contrasts), fMRI images, related features and connectivity data [18], [19].

In addition, the availability of significant computational resources and the advancements of DL algorithms enabled the application of these techniques to improve the accuracy of computer-assisted diagnoses. Also, Deep Learning (DL) models were widely applied for different tasks from both 3D brain volumes, images and connectivity data. As regards models using brain images or volumes, many approaches were investigated, becoming a well-established area of research. In this context, Convolutional Neural Networks (CNNs) have particular importance, since they were recently demonstrated to have remarkable performance in medical analysis, both employing 2D slices and 3D volumes [20]–[22]. In addition, different CNN-based pre-trained models, such as ResNet18, EfficientNet-B0 and VGG etc., were largely employed in neuroimaging research, providing state-of-the-art performance for different tasks [22]–[24]. Subsequently, other approaches, such as Graph Neural Networks (GNNs), were developed. The capabilities of GNNs, which are neural networks that operate on graph data, have recently risen after a decade of development and advancements [25]. These models, along with CNNs and Autoencoders, were adapted and employed to use the brain connectivity matrixes as input.

For example, a connectome-based CNN architecture for classifying Mild Cognitive Impairment was proposed. Two layers of convolution were used, first row by row and then column by column [26]. Similarly, BrainNetCNN is a deeper convolutional network that was also tailored for brain graphs. It uses a CNN-like kernel to compute the convolution of the connections and treats each graph edge as a pixel in an image. The model was validated using structural connectivity matrixes from DTI images for regression task [27]. Subsequently, this model was tested and compared to different ML models with respect to the AD classification task [28]. Alorf and Khan obtained good performance in the AD stages classification employing BC-GCN (Brain Connectivity Graph Convolutional Network), an adapted GNN for brain connectivity data that was previously tested on a regression task [29], and Stacked

Sparse Autoencoders [30]. Also, graph variational autoencoder employing both structural and functional connectivity was employed on AD dataset to find a unified embedding via a classification task [31].

Despite the effectiveness of DL algorithms in various classification tasks, their widespread adoption and trust in clinical settings is limited due to their well-known "black box" nature. The latest advancements in Explainable Artificial Intelligence (XAI) techniques aim to bridge the gap between the performance of DL models and the need for human comprehension of their processes. However, validation on human brain graphs was not widely performed yet, as for T1-weighted, FLAIR, T2-weighted images and volumes.

Among all approaches, the most widespread XAI methods in neuroimaging can be roughly classified into two categories. First, gradient/feature-based methods use gradients or hidden feature maps to determine the importance of different input features in the model's predictions. Examples are Class Activation Mapping (CAM), Grad-CAM with different variations or guided back propagation [32]. They were used in recent imaging studies [21], [33], but also to interpret decision of recent models employing connectivity data [34], [35]. Second, perturbation-based methods evaluate the effect of changes in input information. The importance of input features is measured by monitoring the variation of the model's output for different input perturbations [32]. These methods were widely employed in many studies using images [21], [23], [36] and connectivity data [37].

The lack of interpretability of DL models necessitated to develop ways for a deeper comprehension of the problem at hand. In fact, XAI methodologies may highlight different features of interest for a possible reinforcement of their roles as brain biomarkers of AD. Additionally, the cost of processing for the extraction of the brain connectivity graphs can be evaluated in relation to the different interpretation and accuracy provided by these data in comparison to the use of more conventional brain images and volumes. To the best of our knowledge, no studies compared AI approaches using different sMRI (structural MRI) data to evaluate and address these issues.

In this work, we employed both brain 3D T1-weigthed volumes and structural connectivity data from the OASIS-3 dataset [38] for AD classification task. Two state-of-the-art DL models, as well as the inherent characteristics of the data, were compared in terms of accuracy and explainability. More specifically: i) we evaluated the performance of DL models such as ResNet18 and BC-GCN-SE for AD classification using multiple brain data: 3D T1-weigthed volumes and structural connectivity data, respectively; ii) we employed an XAI method, Grad-CAM, to assess interpretability of the two DL models. The investigation of XAI was made at different levels highlighting morphological and interregional features of interest in agreement to domain-knowledge and possibly reinforcing their roles as brain biomarkers of AD. Consistent and divergent information employed by the two DL models were analyzed highlighting the advantages of the two methods and data and the potential for the development of superior and more trustworthy DL models.

## 2. MATERIALS AND METHODS

### 2.1. Study population

The dataset used in this study is the third release of the Open Access Series of Imaging Studies (OASIS-3), a longitudinal collection of data focused on the effects of normal aging and early-stage AD [38]. The dataset, released in 2019, includes 1098 participants among which 605 cognitively normal adults (i.e., healthy controls – HC) and 493 subjects at various stages of cognitive decline. Each participant underwent both neuroimaging and clinical assessments, that were conducted independently throughout the study.

The final dataset was obtained with the following steps: i) matching the MRI and clinical data by selecting the closest records within a 3-month time span; ii) matching with the dataset of Amodeo and colleagues [31], that extracted the structural connectivity matrixes from OASIS-3; This resulting dataset was composed of 692 sessions relative to 543 participants (age range 42-95 years, mean age 70.06 ± 8.85 years, F:M = 388:304). Each session was associated to a T1-weigthed scan and a structural connectivity matrix.

Patients at each session were eventually differentiated in HC and AD according to the Clinical Dementia Rating (CDR) Scale (available in the dataset), with normal cognitive functions being represented by a CDR equal to 0, and diseased conditions by a score of either 0.5 (very mild impairment), 1 (mild impairment), or 2 (moderate dementia). The same subject could have been assigned to both classes because different CDRs were quantified from different imaging sessions. These data (2.76% of the overall subjects) were retained within the dataset but properly handled during the training/validation process of DL models (see Sec. 2.3 for further details). The final ratio between HC and AD sessions was 557:135.

## 2.2. Data acquisition and data processing

Of all the available T1-weigted scans, 135 were acquired with a 3T Siemens Biograph_mMR scanner, while the remaining 557 with a pair of 3T Siemens TimTrio scanners (Siemens Medical Solutions USA, Inc). Three different imaging sequences were used, as detailed in Table 1.

The imaging scans were initially processed by Amodeo and colleagues [31] to derive the structural connectivity matrixes used to feed the BC-GCN-SE DL model (see Sec. 2.6 for further details). The original T1-weighted volumes were divided in 132 brain-covering regions. Of these, 91 cortical and 15 subcortical parcels were derived from the FSL Harvard-Oxford maximum likelihood cortical atlas (HOA) [39], while the remaining 26 cerebellar parcels were derived from the Automated Anatomical Labelling atlas (AAL) [40]. Henceforth, the combined HOA and AAL atlas will be referred as HOA + AAL. The corresponding DTI volumes were extracted from T1-weigthed scans as detailed in [31]. The combination of the outlined gray matter regions with the DTI white matter fiber tracking resulted in 692 undirected graphs (i.e., positively weighted connectivity matrixes) that underwent a minimal data processing step according to procedures described in [31]. A diagram summarizing the extraction process for structural connectivity matrixes is reported in the orange box of Fig. 2.

Then, all the available T1-weighted scans were pre-processed using the FSL v.6.0 tool [41] to create a suitable dataset for the ResNet18 DL model (see Sec. 2.5 for further details). First, images were skull-stripped and bias field corrected using the *fsl_anat* script (https://fsl.fmrib.ox.ac.uk/fsl/fslwiki/fsl_anat). Then, they were registered to the standard Montreal Neurological Institute template (1x1x1 mm$^2$) by applying the non-linear warp transformation provided by *fsl_anat.* The images field of view was cropped to a dimension of 148x180x144 voxels to focus on brain tissues while the intensity values were normalised using variance scaling. Eventually, the dimensionality was resized to 115x144x118.

**Table 1.** Acquisition details for the three sequences involved in our study.

|  | 3T Siemens Biograph_mMR | 3T Siemens Biograph_mMR | 3T Siemens TimTrio |
|---|---|---|---|
| Sequence | T1 (MPRAGE_GRAPPA2) | T1 (MPRAGE isoWU) | T1 (MPRAGE) |
| TR (ms) | 2300 | 2400 | 2400 |
| TE (ms) | 2.95 | 2.13 | 3.16 |
| Flip angle (degrees) | 8 | 8 | 8 |
| Voxel size (mm$^3$) | 1.20x1.05x1.05 | 1.00x1.00x1.00 | 1.00x1.00x1.00 |
| FOV (mm$^2$) | 176x240 | 176x256 | 176x256 |
| Slices per slab | 256 | 256 | 256 |
| TI (ms) | 900 | 1000 | 1000 |
| Orientation | Sagittal | Sagittal | Sagittal |

*Legend: MPRAGE, Magnetization Prepared Rapid Acquisition Gradient Echo; GRAPPA2, Generalized Autocalibrating Partially Parallel Acquisition version 2; isoWU, isotropic Weighted Undersampling; TR, repetition time; TE, echo time; FOV, field of view; TI, inversion time.*

## 2.3. Training-validation strategy and evaluation

After pre-processing, the dataset was split into training and validation using a stratified 10-fold cross-validation strategy ensuring that each subject was included in only one of the folds. This approach ensures to preserve the ratio between AD and HC sessions.

Having highly imbalanced classes within the dataset, before training the BC-GCN-SE model (see Sec. 2.6 for further details), we applied the synthetic minority over-sampling technique (SMOTE) [42] to the training set. This data augmentation method allows to generate synthetic samples through a linear interpolation of the real neighboring ones, identified through a k-nearest neighbor approach. This methodology has already been applied to connectivity data in previous studies, though with different target tasks [43], [44]. No data augmentation strategy was instead carried out before training the ResNet18 model (see Sec. 2.7 for further details).

For both models, the obtained performances were evaluated by calculating, across all 10 folds, the median and interquartile range of the true positive (TPR) and true negative (TNR) rates and the median and interquartile range of the classification accuracy.

Once the performance was confirmed to be satisfactory through the k-fold cross validation, the final model was trained by performing a new split of the original dataset while maintaining the same proportion of the cross-validation step (i.e., 90% and 10% of the entire dataset) and the same hyperparameters. This was performed to assess the level of explainability of the DL models under evaluation (see Sec 2.6 and 2.7 for further details). This configuration was used to derive the final AD/HC classification for the entire dataset. Results were evaluated in terms of TPR, TNR and classification accuracy while considering the union of both training and validation data.

## 2.4. ResNet18: Deep Learning model for 3D T1-weighted volumes

The recognition of AD from 3D T1-weighted volumes was conducted using Resnet18, a pre-trained DL model that was trained for multi-class classification of images of the ImageNet database [45], which was adapted for 3D inputs [46]. ResNet18 was selected based on previous studies which demonstrated its good performance in AD recognition compared to other pre-trained models [36]. The model is characterized by 18 layers and its general structure includes a 3D convolutional layer and four sets of residual blocks, each containing two 3D convolutional layers, with a shortcut connection that bypasses the convolutional layers and adds the input directly to the output of the second convolutional layer. Transfer learning was employed on the pre-trained model by adding a set of layers consisting of a Global Average Pooling (GAP) layer, two fully connected layers (128 and 32 neurons) with Rectified Linear Unit (ReLU) activation, and a sigmoid activation unit, as depicted at the top of the Fig. 1.

The binary cross-entropy loss function was used for training, with a batch size of 16 and class weighting. The best model was chosen by applying the early stopping criterion to the validation loss, with a patience of 8. The model with the highest Area Under the Curve (AUC) value across training epochs was retained.

## 2.5. BC-GCN-SE: Deep Learning model for structural connectivity graphs

As mentioned in the Introduction, many GCNs were recently proposed with different purposes and in different applications. However, GCN is a node-based DL method based on features related to the different nodes. These kinds of models focus the pattern extraction on nodes and features, learning the messages passing through the connections. This structure was demonstrated to be very suitable for sparse graph-structured data, containing only a few edges connecting nodes. Conversely, the brain connectivity data are known to yield dense graphs. in the structural case, the DTI-derived metrics used after fiber tractography, such as the number of reconstructed streamlines or the fractional anisotropy, result in graphs with a high number of connections, far from being sparse. Also, latest advancements in the fiber tracking and connectivity definition methodologies facilitated the calculation of dense weighted connectomes also for structural connectivity [47]. As a result, both direct and indirect connections between different brain areas result to be crucial for brain communication. For these reasons, common GCNs usually employed with success in other research fields can be unsuitable for the study of structural and functional connectivity. At the same time, an edge-based GCN, the so-called BC-GCN, was recently proposed and specifically employed in functional connectivity tasks [29], [30]. The BC-GCN model with the possible extension of a Squeeze-and-Excitation block (BC-GCN-SE) resulted in performing well in regression and classification tasks. For this reason, we modified this model to be adapted to our AD classification task with structural connectivity data.

BC-GCN-SE is mainly composed of five major units: the graph path convolution (GPC), which allow the extraction of the feature maps, the edge (EP) and node pooling (NP), the Squeeze-and-Excitation (SE), to emphasize or suppress them, and the fully connected block for classification.

As said, the communication between different brain areas is achieved through a combination of direct and indirect connections. To extract significant information from these pathways and represent the multi-order information, GPC layers are utilized. This module represents the counterpart of convolutional layers in CNNs leveraging meaningful characteristics from high-order paths by stacking multiple layers. EP and NP pooling layers are employed to aggregate information from nodes and edges downsampling the feature maps, which are the output of the convolutional layers. EP and NP are the counterpart of pooling layers in typical CNNs. The SE block is constructed as the typical SE model [29], but the convolution is modified to correspond to the GPC outlined above. SE layer is positioned after each of the three GPC layers. Finally, the classification part, composed of fully connected layers, was composed of two fully connected blocks after the NP and before the final sigmoid activation, adapting the network to a binary classification task. The final employed architecture of the BC-GCN-SE model is summarized in the green block of the Fig. 2.

The binary cross-entropy loss function and batch size of 64 were employed during training. The model selection process involved utilizing the early stopping. Whether a minimum validation loss was not achieved within a certain number of epochs, thus stop decreasing, the training was halted. The model with greater AUC value was retained.

## 2.6. Grad-CAM

To perform XAI on the model's predictions, the Grad-CAM method was employed, which is a technique that generates heatmaps to visualize the important regions of an input image that the model uses for classification [48]. Unlike standard CAM, Grad-CAM is a more general approach that does not require any changes to the architecture of the CNN. CAM, on the other hand, requires the removal of the fully connected block of the network and the addition of a GAP layer followed by a single fully connected layer to obtain a relevance heatmap. Using Grad-CAM, it is possible to employ the original CNN architecture, regardless the application. Indeed, the greater flexibility is given by the fact that it is not only limited to classification tasks and allows to obtain relevance heatmaps from any layer of the network, representing features at different granularities [48]. Grad-CAM, thus, produces a heatmap $g \in R^{x \times y}$ to identify the regions of an input image $i \in R^{X \times Y}$, having the greatest influence on the classifier score in support of a given class $c$. In the context of our application, $i$ represents the structural connectivity matrix, where X and Y are the same, referring to the number of rows/columns, or the 3D T1-weighted volume extending what previously reported to the third dimension, with X, Y and Z equal to the 3D voxel position (3D scan $i \in R^{X \times Y \times Z}$; 3D heatmap: $g \in R^{x \times y \times z}$).

The score for class $c$ is denoted by $s_c$, while $a^k \in R^{x \times y}, k = 1, \ldots, K$ represents the activation maps that correspond to the k-th filter of the selected convolutional layer. The Grad-CAM for class $c$ is achieved by computing a weighted average of $a^k, k = 1, \ldots, K$, followed by a Rectified Linear Unit (ReLU) activation function to consider only the positive contributions:

$$g_c = ReLU \sum_k w_c^k a^k \tag{1}$$

Where:

$$w_c^k = \frac{1}{xy} \sum_x \sum_y \frac{\partial s_c}{\partial a^k(x, y)} \tag{2}$$

$w_c^k$ are the average derivatives of $s_c$ with respect to each element $(x, y)$ of the input matrix in the activation $a^k$ and are indicated as *relevance weigths*. The same is achieved for 3D T1-weigthed volumes simply computing the average derivatives with respect to elements $(x, y, z)$ and dividing by $xyz$.

Typically, following the implementation of CAM, the last convolutional layer is chosen. The resulting heatmap $g_c$ generated by Grad-CAM exhibits low resolution due to the architecture of the model. Consequently, it is necessary to up-sample the heatmap to match the input image size, using bicubic interpolation. This procedure facilitates the overlay of Grad-CAM onto the input image, which enhances the interpretation of model decisions. An alternative method to increase the resolution of Grad-CAM involves selecting activations from previous convolutional layers of the network that exhibit higher spatial resolution.

This trade-off between the identification of class-discriminative features with low spatial extent and fine-grained details and thus the resolution of the Grad-CAM was addressed in several works [49]–[51]. Since the goal of this work was to analyze the contribution to the final classification of features of diverse scale (see Par 2.7 for further details) and especially last layer of ResNet18 produced coarse-grained ones due to inherent architecture, we computed the heatmaps at different layers for every session. Such heatmaps were then averaged for each structural connectivity matrix or 3D T1-weighted volume - as performed in [50] and implemented in [52] – according to the following equations:

$$G_c(x, y) = \frac{\sum_{l=1}^{L} \sum_{x=1}^{132} \sum_{y=1}^{132} g_c^l(x, y)}{L} \tag{3}$$

$$G_c(x, y, z) = \frac{\sum_{l=1}^{L} \sum_{x=1}^{115} \sum_{y=1}^{144} \sum_{z=1}^{118} g_c^l(x, y, z)}{L} \tag{4}$$

Where $G_c$ is the resulting mean heatmap and $L$ is the number of layers considered for each model. More specifically, the output layers of the four stages dividing ResNet18 and the three GPC-SE blocks in BC-GCN-SE model were considered.

### 2.7. Explainability assessment

In order to determine the most important information leveraged by the models, we assessed the contributions of each parcel. More specifically, we related the mean heatmaps of AD and HC sessions obtained from both ResNet18 and BC-GCN-SE to the HOA + AAL atlas to assess which brain regions were mostly involved in the classification. This same parcellation allowed for a direct comparison between the level of explainability of the two models. The involved atlas consisted of the combination between 106 regions of the HOA and 26 cerebellar parcels of the AAL. More precisely, it was used to highlight either nodes from the structural connectivity matrices or volumetric regions from the 3D T1-weighted scans. Thus, from the mean heatmap of one session, the contributions highlighted by each connection of each node (i.e., parcel) of connectivity matrixes or in each voxel of each parcel $p$ of 3D T1-weighted scans were averaged to extract a quantity hereafter called relevance value ($RV$):

$$RV_{p,c} = \frac{\sum_{x=1}^{115}\sum_{y=1}^{144}\sum_{z=1}^{118} M_p(x,y,z)G_c(x,y,z)}{\sum_{x=1}^{115}\sum_{y=1}^{144}\sum_{z=1}^{118} M_p(x,y,z)} \quad (5)$$

$$RV_{p,c} = \frac{\sum_{q=1,q\neq p}^{132} G_c(p,q)}{131} \quad (6)$$

where $M_p(x,y,z)$ is a binary mask obtained from the HOA + AAL atlas to define each parcel $p$. A diagram summarizing the XAI steps for 3D T1-weigthed volumes is reported in the bottom part of Fig. 1, whereas for structural connectivity is reported in the yellow block of the Fig. 2.

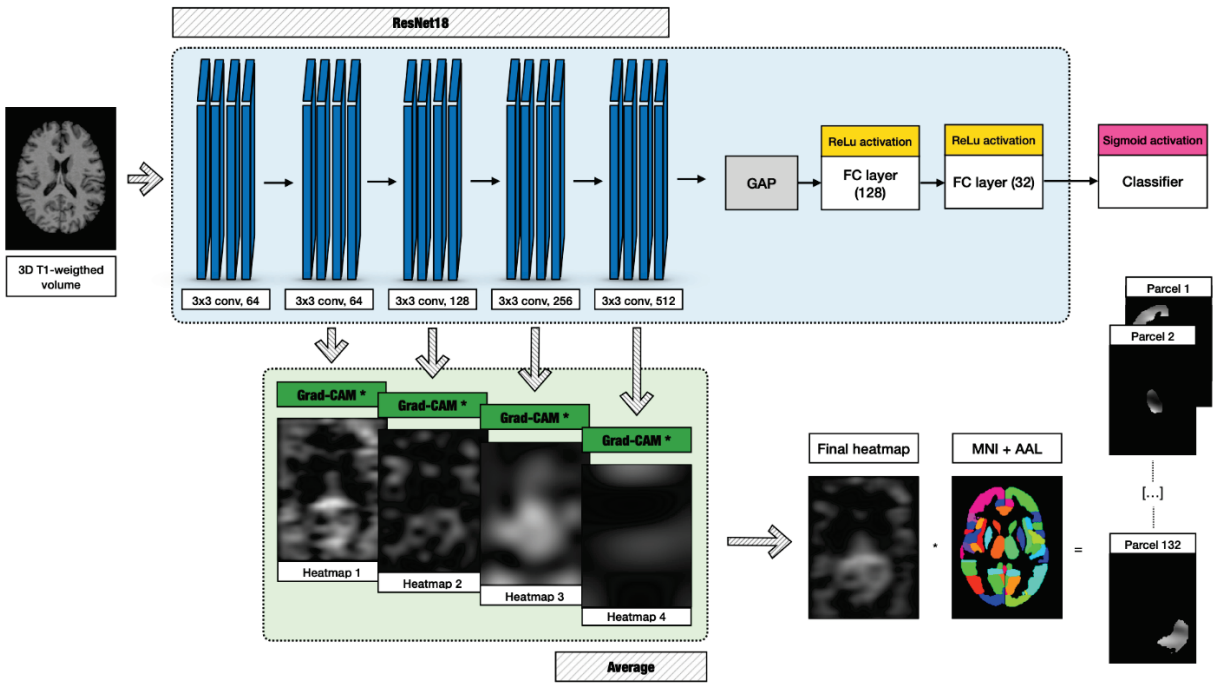

**Figure 1.** Implemented workflow for the AD classification and explainability assessment performed on ResNet18. The architecture of the implemented model (blue panel) comprised 5 convolutional layers, a Global Average Pooling (GAP) layer, three Fully Connected (FC) dense layers with Rectified Linear Unit (ReLU) activation and a sigmoid activation function for the binary classification. The outputs derived from the last 4 convolutional layers were processed using Grad-CAM (green panel) and then averaged. The final heatmap was then multiplied for the binary masks underlying the HOA + AAL atlas parcels.

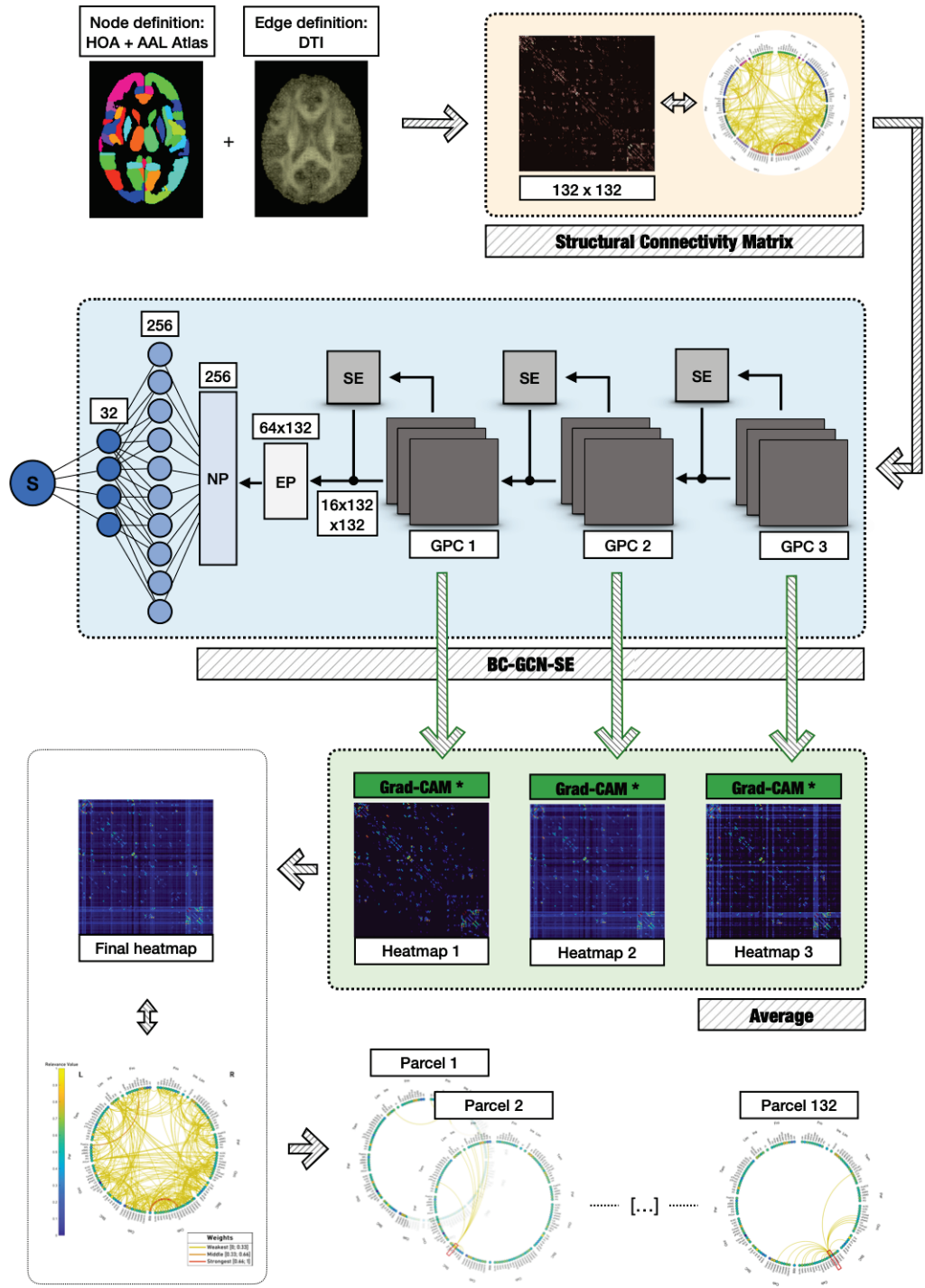

**Figure 2.** Implemented workflow for the connectivity data extraction, AD classification and explainability assessment performed on BC-GCN-SE. The processing steps used to derive the structural connectivity data are displayed in the orange panel. The architecture of the implemented model (blue panel) comprised three Graph Path Convolution layers (GPC), an Edge Pooling (EP) layer, a Node Pooling (NP) layers and two fully connected layers. The outputs derived from the 3 convolutional layers of the model were processed using Grad-CAM (green panel) and then averaged. From the final heatmap the contributions highlighted by each connection of each node of the HOA + AAL atlas were averaged to extract the RV measure of each parcel. A connectogram representing both connectivity edges of the heatmap and color-coded RV values (circle perimeter) is displayed. These plots were created using the SPIDER-NET tool [53].

## 2.7. Statistical Evaluation

### 2.7.1 Anatomical targets

After evaluating the DL models (see Sec. 2.3 for further details) results of the XAI assessment were analyzed in comparison to domain knowledge.

As for the BC-GCN-SE model, we selected as structural connectivity target a hallmark of AD: the DMN [7], [10], [11]. The DMN was defined by 31 parcels of the HOA according to the indications provided in the "Cortical & White Matter Parcellation Manual" (https://cma.mgh.harvard.edu/) and to different previous studies [54]–[57]. Specifically, we defined the DMN according to previous studies highlighting the regions characterizing the network. The medial prefrontal cortex, the posterior cingulate cortex, a portion of the temporal lobe, the precuneus, and the inferior parietal areas like the angular and supramarginal gyri often make up these brain regions. In addition to these usual core regions, the network frequently include the lateral temporal cortex, hippocampus, and amygdala [54], [55], [58]. In this regard, no unique definition exists. While prior research offers preliminary indications of a more comprehensive characterization of the DMN system, additional studies are required to define the anatomical scope of certain parcel contribution [58].

As for the ResNet18 model, a different cluster of parcels was selected as anatomical target. According to previous studies [59], [60], patterns of atrophy in the medial temporal lobe (MTL) represent a well-established sMRI biomarker for AD and is often used a diagnostic criterion for individuals displaying early symptoms. Thus, while conducting the XAI assessment, 4 structures were considered bilaterally, for a total of 8 regions of interest: hippocampus, anterior and posterior parahippocampal gyri and amygdala.

### 2.7.2 Statistical tests

The investigation of the processes underlying the decision of the classifiers was initially carried out by performing a statistical test to identify the different contributions to AD/HC classification of the parcels taken from the HOA + AAL atlas for both models. To predispose the statistical comparison, some operations were carried out. First, we obtained two sets from the RV associated to each parcel of each subject. These sets were created by: (i) removing the misclassified sessions; (ii) aggregating the RV values across single parcels, for subjects with multiple MRI sessions and same class, (iii) Removing the sessions associated to the same subjects but with different classes (removing records); (iv) separating the HC and AD sessions; (v) normalizing between 0 and 1 using the minimum and the maximum values among all RV values obtained from the corresponding models; for example, all RV values obtained with ResNet18 were normalized according to the maximum and minimum obtained from all RV related to each parcel and each correctly classified subject.

The resulting sets were independent and of different size according to the class and the number of sessions misclassified by ResNet18 and BC-GCN-SE.

In order to evaluate the regions and inherent characteristics that were particularly used among all parcels for the classification task of AD, two analyses were performed. First, the Kolmogorov-Smirnov test was performed on all the parcel samples to evaluate normality. The two sets were compared between each other through the Mann-Whitney or independent samples t-test using the Benjamini-Yekutieli correction in both cases [61]. The significance level was set to 0.05 and the correction was applied on the result to adjust p-values accounting for multiple comparisons. In this way, it is possible to obtain the parcels that were used differently by the algorithms to classify one or the other class in relation to all parcels.

Second, most relevant parcels for the classification of AD and HC were investigated separately. These subsets were obtained according to the 15[th] percentile of the highest RV for both classes (20 parcels out of the 132 total HOA + AAL parcels) and both models were analyzed. Indeed, these most relevant parcels are the most employed by the models for the classification.

From these subsets (i.e., most relevant parcels), we analyzed the following subgroups: i) parcels that were also statistically significant. These were strongly influencing the model while also providing a greater contribution towards one of the two classes; ii) parcels that were not statistically significant. Of these, particular attention was paid to the ones in common between the AD and HC groups as they were strongly influencing the classification of both classes without leveraging one.

## 3. RESULTS

The results of the 10-fold cross-validation from the DL models ResNet18 using 3D T1-weighted volumes and BC-GCN-SE using structural connectivity data are summarized in Table 1.

**Table 1.** Normalized confusion matrix for the ResNet18 and BC-GCN-SE model performance, reported as median value and interquartile range.

| Cross-Validation Normalized Confusion Matrix | | Predicted | |
|---|---|---|---|
| **ResNet18** | | P | N |
| **Actual** | P | 0.817 [0.773, 0.846] | 0.183 [0.154, 0.227] |
| | N | 0.183 [0.167, 0.233] | 0.816 [0.767, 0.833] |
| **BC-GCN-SE** | | P | N |
| **Actual** | P | 0.703 [0.672, 0.769] | 0.297 [0.231, 0.328] |
| | N | 0.261 [0.242, 0.302] | 0.739 [0.698, 0.758] |

Both models achieved good performance. Specifically, ResNet18 achieved balanced performance for AD and HC during cross-validation ($TPR_{median}$ = 0.817; $TPR_{IQR}$ = 0.073; $TNR_{median}$ = 0.816; $TNR_{IQR}$ = 0.066). The median accuracy was 0.811 and the interquartile range was 0.073. On the new split, the performance of ResNet18, obtained merging both the training and validation sets, were TPR = 0.985 and TNR = 0.989, with an accuracy of 0.988.

BC-GCN-SE provided slightly inferior performance with respect to ResNet18 ($TPR_{median}$ = 0.703; $TNR_{median}$ = 0.738). More precisely, the model showed slightly higher results and also less variability for HC during cross-validation ($TPR_{IQR}$ = 0.097; $TNR_{IQR}$ = 0.060). The median accuracy is 0.742 and the interquartile range is 0.058. Finally, the results of BC-GCN-SE on training and validation sets on the new split highlighted TPR = 0.956 and TNR = 0.788, with a total accuracy of 0.821.

**Table 2.** Brain parcels displaying a significant difference between the AD and HC level of relevance.

| ResNet18 (3D T1-weigthed volumes) | | | | | | | | BC-GCN-SE (Structural connectivity) | | | | |
|---|---|---|---|---|---|---|---|---|---|---|---|---|
| N. | Parcel* | Adjusted p | N. | Parcel | Adjusted p | N. | Parcel | Adjusted p | N. | Parcel* | Adjusted p | N. | Parcel | Adjusted p |
| 1 | AG_r | < 0.001 | 23 | AC | < 0.001 | 47 | Pal_l | 0.006 | 1 | AG_r | 0.025 | 23 | NAcc_r | 0.015 |
| 2 | aITG_r | < 0.001 | 24 | Amg_l | < 0.001 | 48 | PostCG_l | 0.003 | 2 | aITG_r | < 0.001 | 24 | aMTG_l | < 0.001 |
| 3 | Amg_r | < 0.001 | 25 | CaN_l | < 0.001 | 49 | PrCun | < 0.001 | 3 | Amg_r | 0.017 | 25 | aSMG_r | 0.031 |

| 4 | aMTG_r | 0.002 | 26 | Cereb1_l | < 0.001 | 50 | PreCG_r | 0.024 | 4 | aMTG_r | < 0.001 | 26 | aSTG_l | 0.038 |
|---|---|---|---|---|---|---|---|---|---|---|---|---|---|---|
| 5 | aTFusC_r | < 0.001 | 27 | Cereb6_l | < 0.001 | 51 | SFG_r | 0.010 | 5 | aTFusC_r | 0.014 | 27 | Cereb1_r | 0.028 |
| 6 | BSt | 0.024 | 28 | Cereb6_r | 0.001 | 52 | SubCalC | < 0.001 | 6 | BSt | < 0.001 | 28 | HG_r | 0.016 |
| 7 | Cereb2_r | < 0.001 | 29 | Cereb7_l | < 0.001 | 53 | TOFusC_l | < 0.001 | 7 | Cereb2_r | 0.034 | 29 | ICC_r | 0.031 |
| 8 | Cuneal_r | < 0.001 | 30 | Cereb8_l | < 0.001 | 54 | Tha_r | < 0.001 | 8 | Cuneal_r | 0.016 | 30 | iLOC_r | < 0.001 |
| 9 | FO_r | < 0.001 | 31 | Cereb8_r | < 0.001 | 55 | Ver10 | < 0.001 | 9 | FO_r | 0.001 | 31 | LG_r | 0.044 |
| 10 | FP_r | < 0.001 | 32 | Cereb9_l | < 0.001 | 56 | Ver45 | 0.005 | 10 | FP_r | 0.018 | 32 | OFusG_r | 0.018 |
| 11 | IFG_tri_l | < 0.001 | 33 | Cereb9_r | < 0.001 | 57 | Ver8 | < 0.001 | 11 | IFG_tri_l | 0.018 | 33 | Pal_r | 0.028 |
| 12 | IFG_tri_r | < 0.001 | 34 | Cuneal_l | < 0.001 | 58 | aITG_l | < 0.001 | 12 | IFG_tri_r | 0.044 | 34 | pMTG_r | 0.018 |
| 13 | MedFC | < 0.001 | 35 | FO_l | < 0.001 | 59 | aPaHC_l | < 0.001 | 13 | MedFC | 0.006 | 35 | PO_l | 0.034 |
| 14 | PaCiG_l | < 0.001 | 36 | FP_l | < 0.001 | 60 | aPaHC_r | < 0.001 | 14 | PaCiG_l | < 0.001 | 36 | PO_r | 0.021 |
| 15 | PaCiG_r | < 0.001 | 37 | Hip_l | < 0.001 | 61 | aTFusC_l | < 0.001 | 15 | PaCiG_r | 0.020 | 37 | PostCG_r | 0.010 |
| 16 | PreCG_l | < 0.001 | 38 | Hip_r | < 0.001 | 62 | pITG_l | 0.006 | 16 | PreCG_l | 0.024 | 38 | pSMG_l | 0.032 |
| 17 | SFG_l | < 0.001 | 39 | IC_r | < 0.001 | 63 | pITG_r | < 0.001 | 17 | SFG_l | < 0.001 | 39 | pSMG_r | < 0.001 |
| 18 | Tha_l | < 0.001 | 40 | IFG_oper_l | < 0.001 | 64 | pMTG_l | 0.011 | 18 | Tha_l | 0.031 | 40 | pSTG_r | 0.028 |
| 19 | toMTG_r | 0.045 | 41 | IFG_oper_r | < 0.001 | 65 | pPaHC_r | 0.009 | 19 | toMTG_r | 0.044 | 41 | PT_r | 0.019 |
| 20 | Ver9_r | < 0.001 | 42 | MidFG_l | < 0.001 | 66 | pSTG_l | 0.010 | 20 | Ver9 | 0.024 | 42 | SMA_l | 0.018 |
| 21 | NAcc_l | < 0.001 | 43 | MidFG_r | < 0.001 | 67 | pTFusC_r | < 0.001 | 21 | NAcc_l | 0.003 | 43 | SPL_r | 0.018 |
| 22 | CaN_r | < 0.001 | 44 | OFusG_l | 0.026 | 68 | sLOC_l | < 0.001 | 22 | CaN_r | 0.007 | 44 | toITG_r | < 0.001 |
|  |  |  | 45 | OP_l | < 0.001 | 69 | sLOC_r | < 0.001 |  |  |  | 45 | Ver12 | 0.044 |
|  |  |  | 46 | OP_r | < 0.001 | 70 | toITG_l | < 0.001 |  |  |  | 46 | Ver7 | 0.018 |

*Results relative to the Mann-Whitney or independent samples t-tests are reported in terms of p-value, after performing the Benjamini-Yekutieli correction. The first column of the ResNet18 and BC-GCN-SE model (from N.1 to N.22) is marked with an asterisk (i.e., Parcel\*), to highlight the significant parcels common to both.*

The results from the statistical tests on the parcel XAI heatmaps are summarized in Table 2. Among all the parcels of the atlas used, the relevance values of 70 and 46 parcels of AD in comparison to HC for ResNet18 and BC-GCN-SE respectively were found to be statistically significant ($p < 0.05$). In particular, 22 parcels were in agreement between the two methods (first column of Table 2, highlighted with asterisk), mainly belonging to cortical parcels (except two cerebellar parcels). More specifically, in the case of ResNet18, 7 out of 8 total target MTL and hippocampus parcels, with the only exception of the anterior division of the left parahippocampal gyrus were found to be significant. In the case of BC-GCN-SE, instead, all the parcels belonging to target DMN except the frontal orbital cortex, the posterior part of the cingulate gyrus, the temporal pole, the precuneus and the hippocampus (12 out of 17 parcels) were found to be statistically significant ($p < 0.05$) in at least one hemisphere (70.59%). Considering lateralization, the number was 16 out of 31 total DMN parcels (51.61%). The distributions of the relevance values from both models of these target parcels are shown in Fig. 3 and Fig. 4.

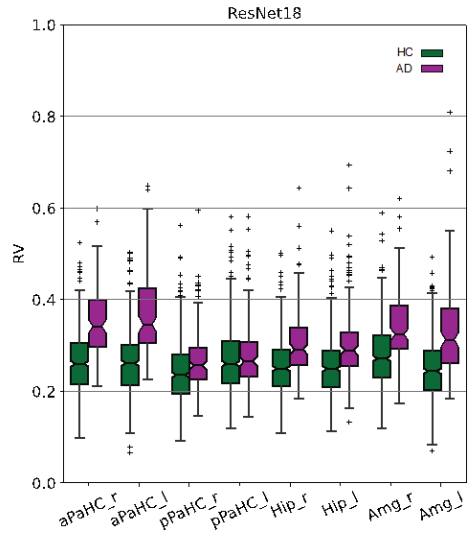

**Figure 3.** Boxplots of the RV values relative to the target parcels (specified on the x axis) for both the BC-GCN-SE model. Results for the AD (in purple) and HC (in green) subjects are resented using different colors.

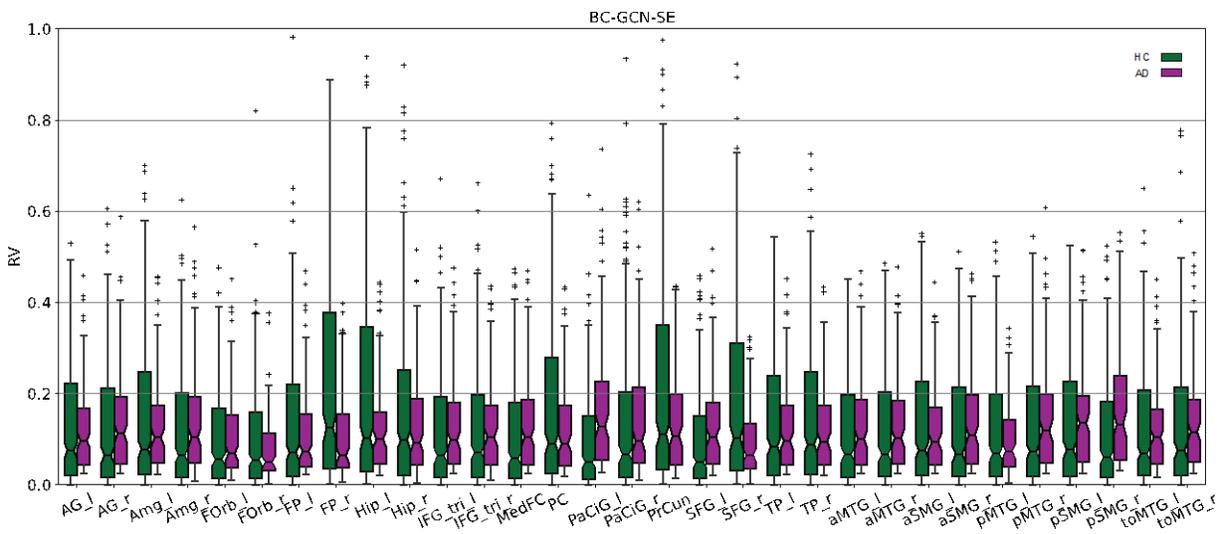

**Figure 4.** Boxplots of the RV values relative to the target parcels (specified on the x axis) for both the BC-GCN-SE model. Results for the AD (in purple) and HC (in green) subjects are resented using different colors.

It is worth noting that most of the regions, with the exceptions of *pPaHC_l* for the ResNet18 model and *FP_r* for the BC-GCN-SE model, resulted to be significantly different with greater relevance values for the AD case.
Afterwards, the most relevant parcels for the classification of AD from both models were analyzed. First, it was noted that 11 out the 20 most relevant parcels for ResNet18 (*Ver10*, *aTFusC_l*, *aTFusC_r*, *aITG_r*, *SubCalC*, *aITG_l*, *aPaHC_l*, *pITG_r*, *MedFC*, *pSTG_l*, *aPaHC_r*), were also found to be statistically significant. The anterior division of both right and left parahippocampal gyrus from target MTL were among these 11 parcels. The remaining 9 most relevant parcels that were not statistically significant were *Cereb10_l*, *Cereb10_r*, *FOrb_l*, *FOrb_r*, *CO_r*, *pSTG_r*, *PO*, *Ver12*, and *Ver3* whose first 7 were also found to be among the most relevant for the HC classification.

Second, 14 out of the 20 most relevant parcels for BC-GCN-SE (*Cereb1_r, iLOC_r, pMTG_r, pSMG_l, pSMG_r, pSTG_r, toITG_r, ICC_r, OFusG_r, PaCiG_l, PaCiG_r, SPL_r, Tha_l, Ver9*), were also found to be statistically different. Of these 14, 5 parcels belong to target DMN. The remaining 6 most relevant parcels that were not statistically significant were *Cereb8_r, Cereb9_l, Cereb9_r, PreCG_r, Ver10, Cereb10_l*, whose first 4 were also found to be among the most relevant for the classification of HC.

To summarize, a diagram indicating the mean RV of all correctly classified sessions of both classes and models per each parcel and divided in lobes are shown in Fig. 5.

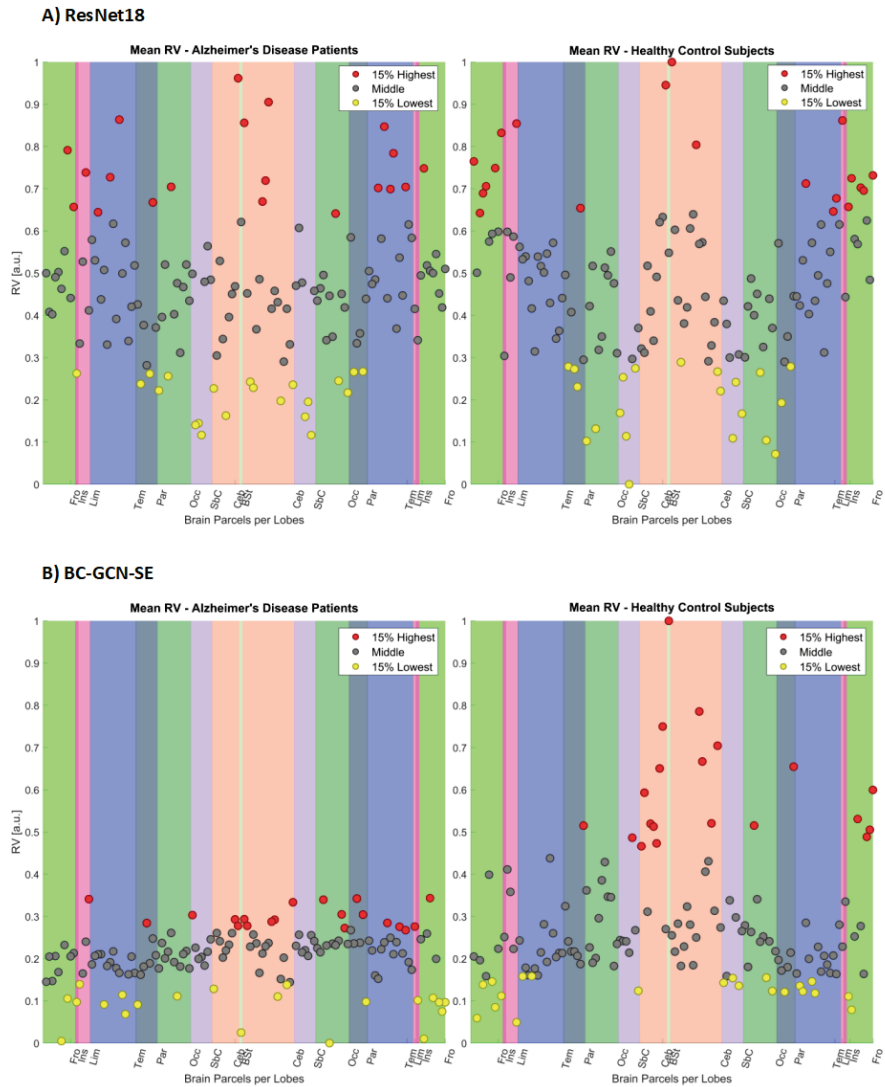

**Figure 5.** Mean RV of all parcels, for the ResNet18 (panel A) and BC-GCN-SE (panel B) models. The AD and HC classes are reported separately. The lobe's belonging is indicated through rectangle colors. The rectangles have different size according to the number of parcels contained in the lobes. The RV of each parcel is labeled by dot colors according to the following criterion: red indicates the 15% of parcels characterized by highest RV, yellow indicates the 15% of parcels characterized by lowest RV, grey indicates the remaining parcels. Plots were created using the SPIDER-NET tool [53]. Fro: Frontal Lobe; Ins: Insular Cortex; Lim: Limbic Lobe; Tem: Temporal Lobe; Par: Parietal Lobe; Occ: Occipital Lobe; SbC: Subcortical Structures; Ceb: Cerebellum; BSt: Brainstem.

## 4. DISCUSSION

In this work, we compared two AI methods trained on different types of sMRI data, namely 3D T1-weighted scans and structural connectivity matrixes extracted from dMRI data. Both approaches were applied to an AD classification task and evaluated first in terms of model accuracy and then in terms of explainability. The population considered was a subset of the OASIS-3 dataset. In particular, we focused on 543 subjects comprising both healthy controls and individuals at various stages of AD cognitive decline. Since a group of subjects underwent multiple imaging sessions that we treated as separate records, the final dataset consisted of 692 MRI volumes and structural connectivity matrixes, with an HC:AD ratio of 557:135.

### 4.1 DL model performance

Overall, the DL models highlighted good classification results. On the one hand, the BC-GCN-SE obtained acceptable performance with respect to the existing literature [28]. Higher accuracy ($TNR_{median}$ = 0.738) and slightly lower cross-validation variability ($TNR_{IQR}$ = 0.060) was found for the HC classification as compared to the AD case ($TPR_{median}$ = 0.703; $TPR_{IQR}$ = 0.097). On the other hand, the ResNet18 model achieved slightly superior accuracies with respect to the BC-GCN-SE when classifying both AD and HC subjects ($TPR_{median}$ = 0.817; $TNR_{median}$ = 0.816). The obtained results were in line with the existing literature [36]. The AD and HC variabilities seemed comparable ($TPR_{IQR}$ = 0.073; $TNR_{IQR}$ = 0.066). To the best of our knowledge, no other studies compared DL models trained using 3D T1-weighted volumes and structural connectivity data neither in terms of accuracy nor explainability. Moreover, this is the first work testing BC-GCN-SE model on structural connectivity with the aim of classifying diverse AD subjects of the OASIS-3 dataset. Few other AD classification studies were conducted on alternative populations, such as the well-known ADNI dataset [28], [30], highlighting good performance for the DL models adapted to graph-structured data - mostly functional connectivity. However, these models are relatively recent, especially when compared to more mature CNNs architectures working on images such as the ResNet18 model. This latter is a widely tested pre-trained model with weights obtained from more than a million images from the ImageNet database, probably resulting in a better ability to generalize.

In addition, it is worth noting that the structural connectivity data obtained from DTI have inherent limitations related to the processing pipelines that may result in noisy connections. The absence of gold-standard methods for the creation of connectivity graphs and for edge weighting is indeed one of connectomics' current biggest shortcomings [62], [63]. DTI fiber tracking algorithms suffer, for example, from the assumption of a single fiber orientation per voxel, resulting in systematic errors for complex fiber geometries (such as crossing, kissing, twisting fibers etc.). In addition, other issues can strongly affect the final extracted data since they are directly dependent on the parameters of the dMRI experiment, and the accuracy can decrease with factors such as pathway length, shape and size of the reference region, and shape of the tract in question [62]. Usually, thresholding methods are widely applied on the raw connectivity matrixes to remove spurious edges, although the optimal approach is yet to be found. This results in altered brain graph structure and connectedness. Future studies could fruitfully explore this issue further by applying some of these most recent methods to assess the performance after connectivity data cleaning [64].

Given this premises, both results can be considered promising, even considering the great variability of severities of the AD subjects, sessions parameters, the presence of multiple acquisition sessions without a predefined design setting (such as scheduled acquisitions) in the OASIS-3 dataset. Moreover, the division into classes is made according to the CDR values, also resulting in 15 subjects having sessions with different labels and in a majority of sessions from subjects with very mild impairment (n. 97 sessions with CDR 0.5 out of 135 total AD sessions). Clearly, being the range of dementia effects wide and of difficult definition, it results in a more difficult task.

As said, the use of DL models on brain connectivity graphs is quite recent, thus assessing them in comparison to more established approaches would be of great importance. The processing effort of extracting connectivity data, providing new information and relevant data would be thus enhanced if casting new light on the exploration of brain biomarkers and computational cost in parallel to optimal results of AI. Future research should further develop and confirm these first findings, aiming to improve the performance since demonstrated higher potential. This would be of great relevance in context of the definition of prodromal symptoms of the pathologies, diagnosis and rehabilitation.

### 4.2 ResNet18 explainability results

After using the DL models to classify our data, this work focused on their explainability. The aim was to validate the functioning of AI methodologies employing MRI volumes and structural connectivity matrixes and to assess their agreement with respect to domain knowledge. As we applied Grad-CAM on both classification models, results were evaluated across 132 parcels derived from the HOA + AAL brain atlas. At first, we extracted the levels of relevance characterizing each parcel in both the AD and HC subjects and compared results, to determine if the regional information was used primarily to identify either one of the two classes. Then, we investigated the most relevant parcels for the AD classification task according to a ranking criterion applied on the RV measures.

As for the imaging classification task, we saw a significant difference between AD and HC in 70 brain areas. Among these, particular attention was paid to the regions deemed as more relevant to identify neurodegenerative processes from sMRI data, as indicated by the existing literature. Numerous studies highlighted the involvement of the MTL in the pathogenesis of AD, and pointed at its volumetric loss as an early sign of the disease progression [2], [65], [66] . Additionally, the medial temporal atrophy has been indicated among the diagnostic criteria for AD by a revised version of the traditional National Institute of Neurological and Communicative Disorders and Stroke–Alzheimer's Disease and related Disorders association criteria [60]. This version removed the original requirement for dementia onset to classify subjects as AD, thus making the MTL a reliable biomarker to detect the disease presence even before the disability phase occurs (i.e., dementia onset and progression). As mentioned in the Methods section (see Sec. 2.7.1 for further details), the brain structures involved in the MTL are the hippocampus, amygdala and parahippocampal regions, that are all key for the episodic and spatial memory [59]. However, among these, the most validated and established sMRI biomarker for AD is the hippocampus. A number of studies have linked its volumetric loss to the memory decline stages [2], [67], [68]. Additionally, Hall and colleagues proved that the amount of time required by cognitively normal subjects to develop dementia was shorter in the presence of hippocampal atrophy [68]. Based on these considerations we assessed the difference between the amount of repeatability characterising AD and HC subjects in the following parcels: *Hip_r*, *Hip_l*, *Amg_r*, *Amg_l*, *aPaHC_r*, *aPaHC_l*, *pPaHC_r*, *pPaHC_l*. The presence of a significant difference between the AD and HC groups in 7 regions – all except the pPaHC – suggests that the algorithm is leveraging them to perform the classification. Additionally, the higher levels of relevance of the AD class as compared to HC in all 7 structures (see Fig. 3), indicates that the algorithm is mainly utilizing them to identify the presence of AD, rather than a physiological condition. From this we may infer a good level of explainability for the implemented ResNet18 model, that seems to be positively influenced by almost all the anatomical regions that are usually involved in the AD progression process. However, one may argue that the number of parcels displaying a difference in the AD and HC levels of RV is high when compared to the total number of parcels we investigated. This may increase the possibility to detect the target structures belonging to the MTL. To overcome this limitation, we investigated which parcels – among the significant ones were also included in the 20 most relevant (i.e., parcels with the highest RV values) for the AD case. We identified a total of 11 parcels, 2 of which belonged to the MTL: *aPaHC_r*, *aPaHC_l*. This serves as a strong confirmation that the algorithm is leveraging them to identify the presence of AD. Additionally, it is noticeable that almost all the remaining regions – the only exception being represented by Vermis10, a parcel of the human cerebellum – are part of the brain cerebral cortex, that is known to be involved in the process of disease progression [66]. In particular, 6 parcels belonged to the temporal lobe (see Fig. 5), whose atrophy pattern has been linked to neuronal loss, visuospatial, language and behavioral impairment by a number of studies [60], [69], [70]. Furthermore, the presence of *MedFC* and *SubCalC* – belonging to the frontal and limbic cortex, respectively – among the most relevant features could be explained by the fact that, during the latest stages of the disease, atrophy spreads to the remaining cortical areas, sparing only the visual and sensory motor cortex, as indicated by Eskildsen and colleagues [2]. Overall, the obtained results indicate that the designed classification approach is correctly using the structural information provided by the 3D T1-weigthed images to recognize diseased individuals. Even though having all of the 8 target regions among the most relevant ones would have proven the achievement of higher levels of explainability, the outlined findings seem promising and open the way for further investigations on the processes underlying the AD classification. Additionally, it is worth noting that the 11 parcels discussed so far – besides aiding the classification – seem also able to avoid the introduction of potential biases or confounding factors. Indeed, only 2 out of them were among the most explainable features for both AD and HC, while the remaining ones were favoring the sole pathological class, thus proving their role as an sMRI biomarker for AD. However, in this regard, it is important to mention that 7 among the remaining AD most explainable features had very high and comparable RV measures with respect to the HC class. Among these, we found 2 cerebellar and 5 cortical regions (either frontal, temporal or parietal). As for the cortical regions we may speculate that the DL model is using them to extract opposite information with respect to the ones generally used to classify AD. The dichotomy between atrophic and physiological cortical volume could play a significant role in the AD/HC distinction. This consideration could still hold for the cerebellar parcels, since the volumetric loss of their molecular and granular layers, despite not being among the most established AD biomarkers, have been linked to the pathology presence by different studies [71], [72]. However, further investigation is certainly needed to confirm the outlined hypothesis or, more generally, to shed light on how these specific regions are used by the DL model to discriminate between healthy and pathological individuals.

**4.3 BC-GCN-SE explainability results**

Regarding the relevance of structural connectivity data and its use within BC-GCN-SE model to classify AD and HC, the DMN was used as target. Indeed, according to different studies [7], [11], [12], changes in the DMN are well-known markers, since age and Alzheimer's pathology can alter the WM and disrupt the DMN normal functioning. In a DTI study was also found that core parcels of the DMN, such as cingulate structure and hippocampus, are even increasingly gaining attention in the AD field since they are also recognized as key structure of the memory system, confirming sMRI findings [7]. We assessed the relevance of these target regions in the task through BC-GCN-SE model (see Sec. 3 for further details), as previously done for the ResNet18 model.

The GRAD-CAM results assessment highlighted 46 parcels which differently contribute to the classification of the two classes with respect to all parcels. Among these, it was found that more than the half of the total parcels belonging to the DMN were statistically significant ($p < 0.05$) with majority of values in favor to AD. It is also worth noting that an even higher percentage of the target parcels (more than 70%) were relevantly used by the model if not considering lateralization. However, we acknowledge that there are considerable discussions and increasing evidence among researchers as to abnormality of topological asymmetry between hemispheric brain WM in AD and Mild Cognitive Impairment [73], [74]. More specifically, when comparing our results to those of other studies, it must be pointed out that all parcels which were found to exhibit asymmetry in the study by Yang and colleagues agreed to the parcels of only one hemisphere used by our model [73]. Examples were angular gyrus and amygdala, belonging to DMN, or inferior temporal gyrus, cuneal cortex, supplementary motor area and lingual gyrus. In addition, several of these parcels were also found to be among the most relevant for the identification of AD. This can be a valuable indication supporting for hemispheric lateralization and aberration possibly due to the long-range connection loss that would be of great interest to be further investigated.

Also, the relevance of the changes within DMN was confirmed by analyzing the statistically significant and at the same time most explainable parcels for AD class. Indeed, among these 14 parcels, 5 belonged to DMN. However, whether on the one hand the results appear to provide an important involvement of the DMN, on the other hand, important parcels such as hippocampus, posterior cingulate cortex or precuneus, were not accounted by the BC-GCN-SE. Beyond that the case of these missing important DMN parcels, also the MTL and, hippocampus included, appeared to exert a small influence on the classification process. Apart from the right amygdala that was found to be statistically significant, other regions such as parahippocampal cortex appeared to not be relevantly used ($p \geq 0.05$). This may represent a limitation in terms of explainability of the model, since some of these regions are also often included in the DMN [7] and MTL was found to be relevant in dMRI studies [14], although less replicated than DMN core areas [75]. At the same time, it is possible to notice from figure Fig. 5 a wide presence of most explainable relevant parcels focused on the right temporal and parietal lobes. In this regard, other studies highlighted general temporal and parietal lobe disruptions contributing to memory impairment [7], [10], [12].

Other particular cases were related to brainstem, the nucleus accumbens and the cerebellum. These regions were not considered as targets, but their analysis appears to be of interest. Indeed, the brainstem was the parcel mostly differing between the two cases in favour to HC, with the highest RV as shown in figure Fig. 5. At this stage of understanding, we believe that this result may highlight an important influence of the corticospinal tract in healthy subject [76]. In addition, the brainstem was found to be associated to apolipoprotein status, resulting in an altered radial diffusivity [77], [78]. Alteration in AD was also found in a study by Nie and colleagues in the accumbens nucleus, that resulted to be statistically different in AD/HC classification made by BC-GCN-SE in both hemispheres [79]. Moreover, analyzing the most explainable relevant parcels that were in common between the two classes it was found that many cerebellum areas aided the model to recognize both classes. This finding can support the hypothesis of a strong involvement of this area in AD that was only recently reported by some studies [80]. Alternatively, studies of whole-brain graph organization revealed subregions of the cerebellum connected with the cortical regions of the DMN [58]. Although the important role of the cerebellum that could be investigated in future works, it is worth noting that the uncertainty of the connectivity data can have a great influence of all these results, especially if referring to cerebellar parts that can be massively affected by noise [81]. In general, these unexpected parcels and less replicated results need further investigation which may potentially bring to new findings regarding the involvement of certain areas in AD or to highlight limitation of the model both in terms of interpretability and accuracy of the results. It is indeed paramount to focus on the interpretation of graph-structured based DL models to provide new evidence of their reliability and trust, and to improve their performance through possible less arbitrary data cleaning and thresholding.

### 4.4 Limitations and perspectives

The present study is not without limitations. For example, different approaches for XAI exist and they may not always be effective in all scenarios [32]. The use and comparison to other methods such as perturbation or distillate methods would be of great interest to further validate the results. In addition, this work only focuses on the mean relevance value extracted from a whole parcel to allow straightforward comparison to 3D T1-weigthed volumes. It would also be appealing, in the case of structural connectivity data, to investigate most explainable connections which could highlight most important long-range connections and their loss in AD [8], [9]. In addition, beyond that the DMN analysis, the investigation of other Resting State Networks (such as Frontoparietal network in relation to fronto-temporal dementia etc.) may highlight new insights on the employed models. Further investigations on the WM might also be performed in the 3D T1-weigthed volumes to visualize similarities and differences with respect to WM metrics related to structural connectivity data. A particular case is related to the presence of White Matter Hyperintensities which can represent important biomarkers of AD condition [82]. In a previous preliminary study, we confirmed their relevance within a DL model employing FLAIR images, although with a low number of subjects [23]. A possible perspective would be to further investigate their effects in regard to connectivity data as well. For example, it would be of great interest to employ tools such as NeMo to extract the effect of these lesions within connectivity data structure [83].

## 4.5 Comparing and combining explanations

The agreement, peculiarities and limitations of the two DL models employing sMRI and dMRI data emerge as key findings. First, 22 regions were found to be significantly used by both models to distinguish the pathological condition. Furthermore, among all the other parcels that did not agree, it was still noticed a large agreement on specific regions without considering hemisphere's belonging. Some other parcels were found to be important, both including the target ones and not. Indeed, although adherence to known markers from 3D T1-weigthed volumes and structural connectivity data are given, other unexpected parcels, mainly from the cortex, resulted in being important, reducing the strength of the explanation. It is worth also noting that the population was composed of subjects at various AD stages, severities, and affections that might simply suggest that different involvement of the disease in different MRI sessions lead to general extensive importance of a considerable portion of the cortex. This is particularly true for the ResNet18 model, that points out more statistically significant parcels. In addition, it is a deeper model with respect to BC-GCN-SE, possibly focusing on more complex features. Together with the comparison between AD and HC most relevant regions were investigated. Among them, the two DL models employing different data highlighted a notable agreement to target. First, it was found that some DMN regions have good importance in the classification from structural connectivity matrixes. Second, it was also shown as the regions characterizing the MTL, found as replicated hallmark in sMRI studies, contain inherent important features for the ResNet18 model employing 3D T1-weighted volumes. In this context, it is also of great interest to underline how some important regions of the MTL such as amygdala and parahippocampal gyri, and hippocampus that were also found to be different in AD connectivity data [14], were not underlined by the XAI analysis of BC-GCN-SE. Indeed, apart from right amygdala, that was found to be statistically significant, all the other regions of the MTL did not result from either statistical test and examining most relevant parcels. This finding may highlight a limitation of BC-GCN-SE model interpretability or the effect of some noise sources inherently present in the connectivity data. At the same time, this complementary relevance of parcels of interest appears promising in the perspective of developing superior and more trustworthy models. Indeed, the use of well-known hallmarks from multiple measures may offer the opportunity of focusing on different information that would be of great interest and significance if used concurrently. In this context, only few studies considered the combination of morphological features of regions from 3D T1-weigthed volumes and interregional properties obtained through structural connectivity data, but that may potentially lead to more accurate results and to a better interpretation [84]. Nowadays, there is indeed the potential to easily collect multiple data from multiple modalities, and in this sense, efforts to assure even greater use of the whole potentiality of DL models, exploiting their peculiarities would be of unvaluable interest for diagnosis and rehabilitation of the pathology. The development of better and more interpretable models can represent accurate and robust solutions to the well-known problem of trust in "black-box models", which limited their diffusion within real settings so far.

## 5. CONCLUSION

In this work, we assessed two DL models working on data from multiple MRI acquisitions performed on a subset of AD and HC subjects from the OASIS-3 dataset. Specifically, we employed 3D T1-weigthed sMRI volumes (ResNet18) and structural connectivity matrixes (BC-GCN-SE) defined according to the HOA + AAL atlas and to dMRI metrics. In this perspective, we compared the models according to their accuracy and explainability. The method employed was GRAD-CAM, which pointed out some target regions, found to highlight markers of the pathology in sMRI and dMRI. More precisely, the expected involvement of the MTL was found using ResNet18, whereas the DMN relevance was found to be important in the decision made by the BC-GCN-SE. Even though an important involvement of these targets in the classification decision, part of these expected regions was missing in the analysis of one of the two models, highlighting complementary explanations. For example, important areas such as hippocampus or parahippocampal gyri were identified by the ResNet18, whereas excluded by BC-GCN-SE. This work emphasized the potential held by imaging and connectivity data as for the creation of better and more reliable models. The opportunity to focus on different information may be provided by well-known hallmarks from multiple measures, if employed concurrently. In this context, combining interregional properties found by brain connectivity graphs with the morphological characteristics of regions from 3D T1-weigthed scans would result in more accurate findings and better interpretation. In summary, the comparison of models using multiple data points highlights the strengths of the two modalities that might help in the creation of more understandable models and, in the long-term, that may result in a rise of the confidence and trust laid in AI models.